\documentclass[final,3p,times,twocolumn,sort&compress]{elsarticle}
\usepackage{amsmath,amssymb,mathtools}
\usepackage{mathrsfs}
\usepackage{tensor}
\usepackage{color,xcolor}
\usepackage[colorlinks]{hyperref}
\AtBeginDocument{\hypersetup{linkcolor=magenta,urlcolor=magenta,citecolor=magenta}}
\allowdisplaybreaks
\makeatletter
\newcommand{\rmnum}[1]{\romannumeral #1}
\newcommand{\Rmnum}[1]{\expandafter\@slowromancap\romannumeral #1@}
\makeatother
\usepackage{changes}
\makeatletter
\@namedef{Changes@AuthorColor}{blue!75!black}
\colorlet{Changes@Color}{blue!75!black}
\makeatother
\begin{document}
\begin{frontmatter}
\title{Complexity factor for a static self-gravitating sphere in Rastall\textendash{}Rainbow gravity}
\author[1]{Zhou-Li Ye}
\author[1]{Yu Wang}
\author[1]{Rui-Xin Yang}
\author[1]{Dao-Jun Liu}\cortext[cor]{Corresponding author}\ead{djliu@shnu.edu.cn}
\address[1]{Department of Physics, Mathematics and Science College, Shanghai Normal University, Shanghai 200234, China}
\begin{abstract}
We generalized Herrera's definition of complexity factor for static spherically symmetric fluid distributions to Rastall-Rainbow theory of gravity. For this purpose, an energy-dependent equation of motion is employed in accordance with the principle of gravity's rainbow. It is found that the complexity factor appears in the orthogonal splitting of the Riemann curvature tensor, and measures the deviation of the value of the active gravitational mass from the simplest system under the combined corrections of Rastall and rainbow. In the low-energy limit, all the results we have obtained reduce to the counterparts of general relativity when the non-conserved parameter is taken to be one. We also demonstrate how to build an anisotropic or isotropic star model using complexity approach. In particular, the vanishing complexity factor condition in Rastall-Rainbow gravity is exactly the same as that derived in general relativity. This fact may imply a deeper geometric foundation for the complexity factor.
\end{abstract}
\end{frontmatter}
\section{Introduction}\label{sec:intro}
Just as the word suggests, complexity is a measure of how complex is a system. Humans seem to be intrinsically endowed with the concept of complexity, which we are bound to feel when a lot of simple things come together by some mechanism and reductionism fails. However, what is really relevant is to evaluate complexity, quantitatively. We need ways to tell precisely how much more complex one system is than another.

So far, complexity does live in many areas, cf., e.g., Refs.~\cite{Kolmogorov,Grassberger1986,Lloyd:1988cn,PhysRevLett.63.105,10.1063/1.2810163,Parisi_1993,LOPEZRUIZ1995321,FELDMAN1998244,PhysRevE.63.066116,PhysRevE.66.011102,SANUDO20085283,PANOS20092343,Susskind:2014moa,Brown:2015lvg}. Even though there is no consensus on a definition yet, underlying almost all of them is a very important quantity---the entropy---and a basic creed that complexity should not only measure the internal structure of a system, but also be in line with one's intuitive perception to a large extent. Perhaps the most typical example was given by L\'opez-Ruiz et al. \cite{LOPEZRUIZ1995321}: the perfect crystal is completely ordered and thus has the lowest entropy and information content, the isolated ideal gas, on the contrary, but both of the simplest models have zero complexity.

For astrophysical background, the statistical complexity of several compact objects has been studied and some illuminating results are obtained \cite{Sanudo:2008bu,Chatzisavvas:2009xu,DeAvellar:2012zz}. Nevertheless, the definition used in these works (based on L\'opez-Ruiz et al. \cite{LOPEZRUIZ1995321} as well as continuous extension thereof \cite{PhysRevE.66.011102}) is only concerned with the energy density of fluid, other variables that describe the equilibrium configuration, the pressure in particular, are not taken into account. Pressure is necessary. It is the relative magnitude of pressure and density that determines the strength of relativistic gravitational effects.

In \cite{Herrera:2018bww}, Herrera proposed a new satisfactory definition of complexity for static spherically symmetric self-gravitating fluid distributions in the context of general relativity (GR). What he called the complexity factor is found in the orthogonal splitting of the Riemann tensor and, specifically, is one of the structure scalars. The new definition contains all components of the energy-momentum tensor and exhibits a minimum value for the simplest system composed of a perfect fluid with homogeneous density and isotropic pressure. Furthermore, it is noteworthy that this definition does not directly relate to entropy or information, at least superficially, but instead takes a proper place for itself in the active gravitational mass. The complexity factor is widely accepted and has been developed in many aspects, such as charged fluid \cite{Sharif:2018pgq}, dissipative case \cite{Herrera:2018czt}, cylindrical and axisymmetric systems \cite{Sharif:2018efi,Herrera:2019cbx}, and modified theories of gravity \cite{Abbas:2018idr,Sharif:2019efn,Sharif:2019sze,Yousaf:2020xux,Yousaf:2020xft,Nazar:2021hgg,Bhatti:2021jue,Bhatti:2021pxr,Yousaf:2021xex,Sharif:2022tbp,Bhatti:2022pqe}.

Although one of the most elegant theories in classical physics, there are many clues revealing that GR is far from being the ultimate answer to gravitational interactions. Rastall gravity is a possible alternative theory of GR, in which geometry and matter are coupled in a non-minimal way \cite{Rastall:1972swe,Rastall:1976uh}. The motivation for modifications is the fact that the covariant conservation law of energy-momentum tensor for a matter field has only been tested in flat spacetime, while it may no longer hold for curved spacetime. Out of respect for the experiments, Rastall assumes that
\begin{equation}\label{eq:non-con}
\nabla^{\mu}T_{\mu\nu}=\lambda\nabla_{\nu}R,
\end{equation}
where $T_{\mu\nu}$ denotes the energy-momentum tensor, $R$ is the curvature scalar and $\lambda$ known as Rastall parameter. The generalized equation of motion, which is compatible with Eq.~\eqref{eq:non-con}, is given by\footnote{We work in geometric units with $c=G=1$.}
\begin{equation}\label{eq:Req}
R_{\mu\nu}-\frac{k}{2}g_{\mu\nu}R=8\pi T_{\mu\nu}
\end{equation}
with $k\equiv 1-16\pi\lambda$, $R_{\mu\nu}$ is the Ricci tensor. Note that when $k=1$, or equivalently $\lambda=0$, Eq.~\eqref{eq:Req} reduces to the Einstein field equations with conserved energy-momentum tensor. The trace of Eq.~\eqref{eq:Req} yields
\begin{equation}
R=\frac{8\pi}{1-2k}T,
\end{equation}
where $T\equiv g^{\mu\nu}T_{\mu\nu}$. Obviously, we must have $k\neq 1/2$. 

It should be honest and pointed out that Rastall gravity is a relatively straightforward but contentious proposition, despite its wide utilization by researchers in both cosmology and astrophysics over the past few decades. Notably, Visser \cite{Visser:2017gpz} claimed that Rastall gravity is completely equivalent to Einstein's GR by rearranging the matter sector into a physical conserved energy-momentum tensor. 
However, Darabi et al. argued that Rastall gravity is an ``open'' theory compared to GR, and these two theories are not equivalent \cite{Darabi:2017coc}.
Actually, the fact that Rastall gravity is not merely an avatar of standard GR is supported by much evidence from both theoretical and observational aspects \cite{Smalley_1983,Al-Rawaf:1995xkt,Maulana:2019evy,Li:2019jkv,Manna:2019onx,Saleem:2020bwa,ElHanafy:2022kjl,Saleem:2024ojw}. Here  we adopt the perspective that Rastall gravity is distinct from GR.

Another theory called ``gravity's rainbow'' was proposed by Magueijo and Smolin in \cite{Magueijo:2002xx}. They extended the nonlinear special relativity (or doubly/deformed special relativity, where the Planck energy $E_{\text{P}}^{}$ serves as a universal constant for all inertial reference frames, instead of the speed of light  \cite{Bruno:2001mw,Kowalski-Glikman:2001vvk,Amelino-Camelia:2000cpa,Magueijo:2001cr,Magueijo:2002am}) to incorporate the effects of gravitational field. As a result, the classical geometry depends on the observer's energy $E$, and the spacetime metric can be expressed in terms of the corresponding tetrad as
\begin{equation}\label{eq:ene-dep}
ds^{2}=\eta_{\mu\nu}\,e^{\mu}(\epsilon)\otimes e^{\nu}(\epsilon)
\end{equation}
with
\begin{equation}\label{eq:cobasis}
e^{0}=\frac{1}{\Xi(\epsilon)}\tilde{e}^{0},\quad e^{i}=\frac{1}{\Theta(\epsilon)}\tilde{e}^{i},
\end{equation}
in which $\eta_{\mu\nu}$ is the Minkowski metric, argument $\epsilon\equiv E/E_{\text{P}}^{}$, $\tilde{e}^{\mu}$ denotes the energy-independent basis, and two rainbow functions are required to satisfy the following conditions
\begin{equation}\label{eq:IR}
\lim_{\epsilon\to 0}\Xi(\epsilon)=1,\quad\lim_{\epsilon\to 0}\Theta(\epsilon)=1.
\end{equation}
Consequently, the quantities derived from the metric, such as curvature and energy-momentum tensor, also become energy dependent. An intriguing fact about rainbow theory is that the location of the event horizon of a black hole remains unchanged, however, the area of the horizon is a function of $\Theta$ \cite{Magueijo:2002xx}.

In Ref.~\cite{Mota:2019zln}, Mota et al. integrated Rastall and rainbow theories in a unified formalism to investigate the internal structure and observational properties of neutron stars. After taking into account Eq.~\eqref{eq:ene-dep}, the original Rastall field equations \eqref{eq:Req} are naturally replaced by a one-parameter family that run over the different energy scales in theoretical space, i.e.,
\begin{equation}\label{eq:eomgra1}
R_{\mu\nu}(\epsilon)-\frac{k}{2}g_{\mu\nu}(\epsilon)R(\epsilon)=8\pi T_{\mu\nu}(\epsilon).
\end{equation}
It has been argued in \cite{Mota:2019zln} that a slight departure from GR is sufficient to have a significant impact on the mass-radius profile of neutron stars. The inclusion of two free parameters is helpful in interpreting astrophysical observations and makes Rastall\textendash{}Rainbow (RR for short) gravity more competitive. In recent years, rich literature has shown an increasing interest in applying RR theory to astrophysical scenarios \cite{Debnath:2019eor,Das:2022vxq,Mota:2022zbq,Li:2023fux,Tangphati:2023nwz,Pradhan:2023vhn,Jyothilakshmi:2023cao,Tangphati:2023fey,Li:2024uwv}. In this work we generalize Herrera's complexity factor to include RR gravity, thereby enhancing our understanding of compact objects within this theoretical framework.

This manuscript is assembled in the following pattern. In the next section, we set the stage for a static self-gravitating fluid sphere with locally anisotropic matter and find out the relations between its Weyl tensor and Tolman mass as a preparation. In Sec.~\ref{sec:orthsprie}, we implement the orthogonal splitting of the Riemann tensor to acquire the structure scalars, one of which would be identified as the complexity factor. Examples of both isotropic and anisotropic solutions are provided in Sec.~\ref{sec:vancom}. Finally, in Sec.~\ref{sec:conclu} a concluding remark is given. Throughout the paper, we choose the metric signature $(+,-,-,-)$ to be consistent with most literature on complexity factor.
\section{A static, spherically symmetric and anisotropic system}\label{sec:ssas}
In this section, we establish the physical variables and equations that determine the internal structure of a non-spinning self-gravitating fluid sphere with locally anisotropic matter. To this end, our starting point is the static, spherically symmetric metric whose general form in the Schwarzschild coordinate system $\{t,r,\theta,\varphi\}$ is given by
\begin{equation}\label{eq:metric0}
ds^{2}=e^{2\alpha(r)}dt^{2}-e^{2\beta(r)}dr^{2}-r^{2}d\theta^{2}-r^{2}\sin^{2}\theta d\varphi^{2}.
\end{equation}
It is easy to verify that
\begin{equation}
\tilde{e}^{0}=e^{\alpha}dt,\ \tilde{e}^{1}=e^{\beta}dr,\ \tilde{e}^{2}=rd\theta,\ \tilde{e}^{3}=r\sin\theta d\varphi
\end{equation}
form a set of orthonormal dual basis. With the help of Eq.~\eqref{eq:ene-dep} and \eqref{eq:cobasis}, we may turn Eq.~\eqref{eq:metric0} into the energy-dependent form
\begin{equation}\label{eq:metric}
ds^{2}=\frac{e^{2\alpha}}{\Xi^{2}}dt^{2}-\frac{e^{2\beta}}{\Theta^{2}}dr^{2}-\frac{r^{2}}{\Theta^{2}}d\theta^{2}-\frac{r^{2}}{\Theta^{2}}\sin^{2}\theta d\varphi^{2}.
\end{equation}
The standard static spherically symmetric line element is recovered in the infrared limit $\epsilon\to 0$, according to Eq.~\eqref{eq:IR}.

On the other hand, we regard the matter within the object to be composed by an anisotropic fluid, its energy-momentum tensor is simply
\begin{equation}\label{eq:enemom}
T_{\mu\nu}=\rho(r)U_{\mu}U_{\nu}-Ph_{\mu\nu}+\Pi_{\mu\nu}
\end{equation}
with
\begin{align}
U_{\mu}=\left(e^{\alpha}/\Xi,0,0,0\right)&,\quad h_{\mu\nu}\equiv g_{\mu\nu}-U_{\mu}U_{\nu},\notag\\
P\equiv\frac{p_{r}^{}(r)+2p_{t}^{}(r)}{3}&,\quad\Pi_{\mu\nu}\equiv\Pi N_{\!\langle\mu\nu\rangle}
\end{align}
in which $\rho$ is the energy density, $\Pi\equiv p_{r}^{}-p_{t}^{}$ characterizes the difference between radial pressure and tangential pressure at each point, $h_{\mu\nu}$ is the induced metric on the hypersurface orthogonal to the fluid velocity $U^{\mu}$, and the symmetric tracefree tensor is defined by
\begin{equation}
N_{\!\langle\mu\nu\rangle}\equiv N_{\mu}N_{\nu}+\frac{1}{3}h_{\mu\nu},
\end{equation}
where the spatial vector $N_{\mu}=(0,-e^{\beta}/\Theta,0,0)$.

Feeding Eqs.~\eqref{eq:metric} and \eqref{eq:enemom} on both sides of the field equations \eqref{eq:eomgra1}, we obtain
\begin{align}
8\pi\rho&=\frac{\Theta^{2}}{e^{2\beta}}\frac{-1+e^{2\beta}+2r\beta^{\prime}}{r^{2}}+\frac{1-k}{2}R,\label{eq:zerozero}\\
8\pi p_{r}^{}&=\frac{\Theta^{2}}{e^{2\beta}}\frac{1-e^{2\beta}+2r\alpha^{\prime}}{r^{2}}-\frac{1-k}{2}R,\label{eq:oneone}\\
8\pi p_{t}^{}&=\frac{\Theta^{2}}{e^{2\beta}}\left(\alpha^{\prime\prime}-\alpha^{\prime}\beta^{\prime}+\alpha^{\prime 2}+\frac{\alpha^{\prime}-\beta^{\prime}}{r}\right)-\frac{1-k}{2}R,\label{eq:twotwo}
\end{align}
where a prime denotes an $r$-derivative and Ricci scalar reads
\begin{equation}
R=\frac{2\Theta^{2}}{e^{2\beta}}\left(\alpha^{\prime\prime}-\alpha^{\prime}\beta^{\prime}+\alpha^{\prime 2}+2\frac{\alpha^{\prime}-\beta^{\prime}}{r}+\frac{1-e^{2\beta}}{r^{2}}\right).
\end{equation}

It is convenient to introduce the mass function as follows\footnote{We label the coordinates by $x^{0}=t$, $x^{1}=r$, $x^{2}=\theta$, $x^{3}=\varphi$.}
\begin{equation}\label{eq:masfun}
\tensor{R}{^3_{232}}=1-e^{-2\beta}\equiv\frac{2m(r)}{r},
\end{equation}
and reformulate the above equation system as:
\begin{align}
m^{\prime}&=\left(4\pi\rho-\frac{1-k}{4}R\right)\frac{r^{2}}{\Theta^{2}},\label{eq:m1r}\\
\alpha^{\prime}&=\frac{\left[4\pi p_{r}^{}+\left(1-k\right)R/4\right]\Theta^{-2}r^{3}+m}{r\left(r-2m\right)},\label{eq:alpha1r}\\
p_{r}^{\prime}&=-\left(\rho+p_{r}^{}\right)\alpha^{\prime}-\frac{2}{r}\Pi-\frac{1-k}{16\pi}R^{\prime}.\label{eq:pr1r}
\end{align}
The last one, the generalized Tolman-Opphenheimer-Volkoff equation derived from Eq.~\eqref{eq:non-con} is equivalent to the 22-component \eqref{eq:twotwo} under the premise that Eqs.~\eqref{eq:zerozero} and \eqref{eq:oneone} hold.

Outside the fluid sphere, the spacetime geometry is described by the modified Schwarzschild solution \cite{Magueijo:2002xx}
\begin{equation}
ds^{2}=\frac{1-2M/r}{\Xi^{2}}dt^{2}-\frac{\left(1-2M/r\right)^{-1}}{\Theta^{2}}dr^{2}-\frac{r^{2}}{\Theta^{2}}d\Omega^{2},
\end{equation}
where $M$ is the total mass of the system and $d\Omega^{2}$ the metric on a unit sphere.

In order to match smoothly the interior and exterior region at body's surface $r=R_{\Sigma}^{}=\text{const}$, we impose the junction conditions of the continuity of induced metric and extrinsic curvature across that surface, which imply
\begin{equation}\label{eq:BC}
e^{2\alpha(R_{\Sigma}^{})}=e^{-2\beta(R_{\Sigma}^{})}=1-\frac{2M}{R_{\Sigma}^{}},\quad p_{r}^{}(R_{\Sigma}^{})=0.
\end{equation}
\subsection{Weyl tensor}
Recall that the Weyl tensor $C_{\rho\sigma\mu\nu}$ is the remainder of the Riemann tensor removing all traces,
\begin{equation}\label{eq:weyldef}
R_{\rho\sigma\mu\nu}=C_{\rho\sigma\mu\nu}+g_{\rho[\mu}R_{\nu]\sigma}-g_{\sigma[\mu}R_{\nu]\rho}-\frac{1}{3}g_{\rho[\mu}g_{\nu]\sigma}R,
\end{equation}
where the square bracket indicates the anti-symmetrized operation over the indices enclosed within.

Similar to the field intensity $F_{\mu\nu}$ in electrodynamics, the Weyl tensor can also be decomposed by an observer into electric and magnetic parts,
\begin{equation}\label{eq:weyldec}
C_{\rho\sigma\mu\nu}=\left(g_{\rho\sigma\alpha\gamma}g_{\mu\nu\beta\tau}-\varepsilon_{\rho\sigma\alpha\gamma}\varepsilon_{\mu\nu\beta\tau}\right)U^{\alpha}U^{\beta}E^{\gamma\tau},
\end{equation}
where $g_{\rho\sigma\alpha\gamma}\equiv 2g_{\rho[\alpha}g_{\gamma]\sigma}$, and $\varepsilon_{\rho\sigma\alpha\gamma}$ being the Levi-Civita tensor, whereas $E_{\mu\nu}\equiv C_{\mu\sigma\nu\rho}U^{\sigma}U^{\rho}$ is a symmetric, tracefree and spatial tensor. The absence of the magnetic part in Eq.~\eqref{eq:weyldec} is attributed to the fact that the Weyl tensor is purely electric for the spherical symmetry, i.e., $B_{\mu\nu}^{}=0$.

Observe that $E_{\mu\nu}^{}$ may also be expressed as
\begin{equation}
E_{\mu\nu}=EN_{\!\langle\mu\nu\rangle}
\end{equation}
with
\begin{equation}\label{eq:elesca}
E=-\frac{\Theta^{2}}{2e^{2\beta}}\left(\alpha^{\prime\prime}-\alpha^{\prime}\beta^{\prime}+\alpha^{\prime 2}-\frac{\alpha^{\prime}-\beta^{\prime}}{r}+\frac{1-e^{2\beta}}{r^{2}}\right).
\end{equation}
Combining Eqs.~\eqref{eq:weyldef}, \eqref{eq:masfun}, \eqref{eq:elesca}, \eqref{eq:eomgra1} and \eqref{eq:enemom}, we find
\begin{equation}\label{eq:masfun1}
\Theta^{2}m=\frac{4}{3}\pi r^{3}\left(\rho-\Pi\right)+\frac{1}{3}r^{3}E-\frac{1-k}{12}r^{3}R,
\end{equation}
from which it is straightforward to obtain
\begin{equation}\label{eq:elesca1}
E=4\pi\Pi-\frac{1}{2r^{3}}\int_{0}^{r}x^{3}\left(8\pi\rho-\frac{1-k}{2}R\right)^{\prime}dx,
\end{equation}
a very useful relationship between the Weyl tensor and density inhomogeneity and pressure anisotropy of the fluid distribution, as well as the RR corrections.

Finally, inserting Eq.~\eqref{eq:elesca1} into \eqref{eq:masfun1} produces
\begin{equation}\label{eq:masfun2}
\Theta^{2}m=\frac{4}{3}\pi r^{3}\rho-\frac{4}{3}\pi\int_{0}^{r}x^{3}\rho^{\prime}dx-\frac{1-k}{4}\int_{0}^{r}x^{2}Rdx.
\end{equation}
It is worth mentioning that this result can also be obtained directly from Eq.~\eqref{eq:m1r}.
\subsection{Tolman mass}
The Tolman mass is a kind of quasi-local energy for static spacetime, which can also effectively measure the total energy of a spherical self-gravitating system \cite{Tolman:1930zz}. In modern notation conventions, the Tolman mass formula is given by \cite{Szabados:2009eka}
\begin{equation}\label{eq:tolmandef}
M_{\text{T}}^{}=\frac{1}{4\pi}\int_{\Sigma}D^{\mu}D_{\mu}f,
\end{equation}
where $\Sigma$ is a spacelike hypersurface enveloping all matter with $U^{\mu}$ as its normal vector, $D_{\mu}$ is the connection compatible with $h_{\mu\nu}$, and $f$ being the redshift of spacetime.

For now, $f=e^{\alpha}/\Xi$, and Eq.~\eqref{eq:tolmandef} becomes
\begin{equation}
M_{\text{T}}^{}=\frac{4\pi}{\Xi\Theta^{3}}\int_{0}^{R_{\Sigma}^{}}r^{2}e^{\alpha+\beta}\left(\rho+3P+\frac{1-k}{8\pi}R\right)dr,
\end{equation}
where Eq.~\eqref{eq:eomgra1} is used. It is useful to localize $M_\text{T}$ as  a function $m_\text{T}$ referred to the mass within a sphere of radius $r\leqslant R_{\Sigma}^{}$,
\begin{equation}\label{eq:tolmanloc}
m_{\text{T}}^{}=\frac{4\pi}{\Xi\Theta^{3}}\int_{0}^{r}x^{2}e^{\alpha+\beta}\left(\rho+3P+\frac{1-k}{8\pi}R\right)dx.
\end{equation}
Using Eqs.~\eqref{eq:zerozero}, \eqref{eq:oneone} and \eqref{eq:twotwo} to finish the integral, we  obtain that
\begin{equation}\label{eq:mTofalpha}
m_{\text{T}}^{}=\Xi^{-1}\Theta^{-1}r^{2}e^{\alpha-\beta}\alpha^{\prime}.
\end{equation}
With the help of the 4-acceleration, $A_{\mu}=-\partial_{\mu}\ln f$, we learn that the magnitude of gravity experienced by a static observer is
\begin{equation}
g=\sqrt{\left\lvert A_{\mu}A^{\mu}\right\rvert}=\Xi\Theta^{2}e^{-\alpha}\frac{m_{\text{T}}^{}}{r^{2}},
\end{equation}
which reveals that the Tolman mass plays the role of active gravitational mass.

Differentiating Eq.~\eqref{eq:mTofalpha} and comparing the result with Eq.~\eqref{eq:elesca} yields \cite{Herrera:1997plx}
\begin{equation}
m_{\text{T}}^{\prime}-\frac{3}{r}m_{\text{T}}^{}=-\Xi^{-1}\Theta^{-3}r^{2}e^{\alpha+\beta}\left(4\pi\Pi+E\right),
\end{equation}
the solution is found to be
\begin{equation}
m_{\text{T}}^{}=m_{\text{T}}^{}(R_{\Sigma}^{})\left(\frac{r}{R_{\Sigma}^{}}\right)^{3}+\frac{r^{3}}{\Xi\Theta^{3}}\int_{r}^{R_{\Sigma}^{}}\frac{e^{\alpha+\beta}}{x}\left(4\pi\Pi+E\right)dx,
\end{equation}
or equivalently,
\begin{align}\label{eq:mTofPiElam}
m_{\text{T}}^{}={}&m_{\text{T}}^{}(R_{\Sigma}^{})\left(\frac{r}{R_{\Sigma}^{}}\right)^{3}+\frac{r^{3}}{\Xi\Theta^{3}}\int_{r}^{R_{\Sigma}^{}}\frac{e^{\alpha+\beta}}{x}\left[8\pi\Pi\phantom{\!\!\!\left(\frac{1}{2}\right)^{\prime}}\right.\notag\\
&\left.\quad-\,\frac{1}{2x^{3}}\int_{0}^{x}y^{3}\left(8\pi\rho-\frac{1-k}{2}R\right)^{\prime}dy\right]dx.
\end{align}
Unlike Eq.~\eqref{eq:masfun2}, the above two expression for Tolman mass properly include the density inhomogeneity and the pressure anisotropy of fluid distribution in the context of RR gravity. As we shall see below, Eq.~\eqref{eq:mTofPiElam} is very helpful in looking for the complexity factor.
\section{Orthogonal splitting of the Riemann tensor}\label{sec:orthsprie}
According to Herrera \cite{Herrera:2018bww}, the complexity factor is hidden in the orthogonal splitting of the Riemann tensor initiated by Bel \cite{bel1961}. We turn, now, to this task.

Consider the following three tensors,
\begin{align}
Y_{\mu\nu}^{}&=R_{\mu\sigma\nu\rho}^{}U^{\sigma}U^{\rho},\\
Z_{\mu\nu}^{}&={}^\ast\!R_{\mu\sigma\nu\rho}^{}U^{\sigma}U^{\rho},\\
X_{\mu\nu}^{}&={}^\ast\!R^{\ast}_{\mu\sigma\nu\rho}U^{\sigma}U^{\rho},
\end{align}
where $\ast$ is called Hodge star, and the left and right dual of the Riemann tensor are, respectively, defined by
\begin{equation}
{}^\ast\!R_{\mu\sigma\nu\rho}^{}\equiv\frac{1}{2}\varepsilon_{\mu\sigma\alpha\gamma}^{}\tensor{R}{^{\alpha\gamma}_{\nu\rho}},\quad R^{\ast}_{\mu\sigma\nu\rho}\equiv\frac{1}{2}\tensor{R}{_{\mu\sigma}^{\alpha\gamma}}\varepsilon_{\alpha\gamma\nu\rho}^{}.
\end{equation}

Using field equations \eqref{eq:eomgra1} with \eqref{eq:enemom} to replace the Ricci curvature in Eq.~\eqref{eq:weyldef}, we may split the Riemann tensor as four terms,
\begin{equation}
\tensor{R}{^{\rho\sigma}_{\mu\nu}}=R^{\rho\sigma}_{(\text{\Rmnum{1}})\mu\nu}+R^{\rho\sigma}_{(\text{\Rmnum{2}})\mu\nu}+R^{\rho\sigma}_{(\text{\Rmnum{3}})\mu\nu}+R^{\rho\sigma}_{(\text{\Rmnum{4}})\mu\nu}
\end{equation}
with
\begin{align}
R^{\rho\sigma}_{(\text{\Rmnum{1}})\mu\nu}={}&16\pi\tensor{\delta}{^{[\rho}_{\![\mu}}U^{\sigma]}U_{\nu]}^{}-16\pi P\tensor{\delta}{^{[\rho}_{\![\mu}}\tensor{h}{^{\sigma]}_{\!\!\nu]}}\notag\\
&\quad-\frac{16}{3}\pi\left(\rho-3P\right)\tensor{\delta}{^{\rho}_{\![\mu}}\tensor{\delta}{^{\sigma}_{\nu]}},\\
R^{\rho\sigma}_{(\text{\Rmnum{2}})\mu\nu}={}&16\pi\tensor{\delta}{^{[\rho}_{\![\mu}}\tensor{\Pi}{^{\sigma]}_{\!\!\nu]}},\\
R^{\rho\sigma}_{(\text{\Rmnum{3}})\mu\nu}={}&4U^{[\rho}U_{[\mu}^{}\tensor{E}{^{\sigma]}_{\!\!\nu]}}-\tensor{\varepsilon}{^{\rho\sigma}_\gamma}\varepsilon_{\mu\nu\tau}^{}E^{\gamma\tau},\\
R^{\rho\sigma}_{(\text{\Rmnum{4}})\mu\nu}={}&\frac{1-k}{3}R\tensor{\delta}{^{\rho}_{\![\mu}}\tensor{\delta}{^{\sigma}_{\nu]}},
\end{align}
where $\varepsilon_{\sigma\mu\nu}\equiv U^{\rho}\varepsilon_{\rho\sigma\mu\nu}$ being the induced volume element associated with $h_{\mu\nu}$.

Then, tensors $X_{\mu\nu}$, $Y_{\mu\nu}$ and $Z_{\mu\nu}$ are easily evaluated by the above equations. The final results are
\begin{align}
X_{\mu\nu}={}&\frac{8}{3}\pi\rho h_{\mu\nu}+4\pi\Pi_{\mu\nu}-E_{\mu\nu}-\frac{1-k}{6}Rh_{\mu\nu},\label{eq:xmunu}\\
Y_{\mu\nu}={}&\frac{4}{3}\pi\left(\rho+3P\right)h_{\mu\nu}+4\pi\Pi_{\mu\nu}+E_{\mu\nu}+\frac{1-k}{6}Rh_{\mu\nu},\label{eq:ymunu}\\
Z_{\mu\nu}={}&0.
\end{align}
By definition, both $X_{\mu\nu}$ and $Y_{\mu\nu}$ are symmetric spatial tensors of type (0,2), they can be written as the sum of their trace and tracefree parts. For example
\begin{equation}
X_{\mu\nu}=X_{\langle\mu\nu\rangle}+\frac{1}{3}\tensor{X}{^\gamma_\gamma}h_{\mu\nu},
\end{equation}
where
\begin{equation}
X_{\langle\mu\nu\rangle}\equiv\tensor{h}{_\mu^\sigma}\tensor{h}{_\nu^\rho}\left(X_{\sigma\rho}-\frac{1}{3}\tensor{X}{^\gamma_\gamma}h_{\sigma\rho}\right).
\end{equation}

Denote the trace $\tensor{X}{^\gamma_\gamma}$ as $X_{\text{T}}^{}$ and the difference between $X_{\langle\mu\nu\rangle}$ and $N_{\!\langle\mu\nu\rangle}$ as $X_{\text{TF}}^{}$, similarly for $Y_{\mu\nu}$ \cite{Herrera:2009zp}. We may extract the four structure scalars from Eqs.~\eqref{eq:xmunu} and \eqref{eq:ymunu} in turn:
\begin{align}
X_{\text{T}}^{}&=8\pi\rho-\frac{1-k}{2}R,\\
X_{\text{TF}}^{}&=\frac{1}{2r^{3}}\int_{0}^{r}x^{3}\left(8\pi\rho-\frac{1-k}{2}R\right)^{\prime}dx,\\
Y_{\text{T}}^{}&=4\pi\left(\rho+3P\right)+\frac{1-k}{2}R,\label{eq:YT}\\
Y_{\text{TF}}^{}&=8\pi\Pi-\frac{1}{2r^{3}}\int_{0}^{r}x^{3}\left(8\pi\rho-\frac{1-k}{2}R\right)^{\prime}dx,\label{eq:YTF}
\end{align}
here Eq.~\eqref{eq:elesca1} has been used. As mentioned before, these structure scalars are candidates for the complexity factor. The question comes down to which one of them is reasonably qualified to be selected as the complexity factor. There are two criteria: (\rmnum1) it should measure the complexity of a system in an appropriate way and include all physical variables that describe the equilibrium configuration; (\rmnum2) it should reduce to Eq.~(54) in Ref.~\cite{Herrera:2018bww} when we set $k=\Theta=\Xi=1$ (GR case), at the very least. From these it is easy to check that only Eq.~\eqref{eq:YTF} is suitable. Therefore, we refer to $Y_{\text{TF}}^{}$ as the complexity factor. Comparing Eq.~\eqref{eq:mTofPiElam} with Eq.~\eqref{eq:YTF}, we get
\begin{equation}
m_{\text{T}}^{}=m_{\text{T}}^{}(R_{\Sigma}^{})\left(\frac{r}{R_{\Sigma}^{}}\right)^{3}+\frac{r^{3}}{\Xi\Theta^{3}}\int_{r}^{R_{\Sigma}^{}}\frac{e^{\alpha+\beta}}{x}Y_{\text{TF}}^{}\,dx,
\end{equation}
which means that $Y_{\text{TF}}^{}$ precisely measures the deviation of the value of the active gravitational mass from the simplest system in the context of RR gravity.

In GR \cite{Herrera:2018bww}, the complexity factor is completely determined by the density inhomogeneity and the pressure anisotropy of fluid distribution, and exhibits a minimum value for the simplest system defined by a fluid with homogeneous density and isotropic pressure (i.e., a Schwarzschild uniform density star). However, it is quickly seen from Eq.~\eqref{eq:YTF} that $Y_{\text{TF}}^{}$ also carries the RR modifications in the present work, and our definition reduces to Herrera \cite{Herrera:2018bww} if the rainbow function $\Theta$ and the dimensionless Rastall parameter $k$ are set to be one, simultaneously. Furthermore, combining Eqs.~\eqref{eq:tolmanloc} and \eqref{eq:YT} recognizes the physical meaning of $Y_{\text{T}}^{}$,
\begin{equation}
m_{\text{T}}^{}=\frac{1}{\Xi\Theta^{3}}\int_{0}^{r}x^{2}e^{\alpha+\beta}Y_{\text{T}}^{}\,dx,
\end{equation}
which implies that $Y_{\text{T}}^{}$ is proportional to the active gravitational mass density of anisotropic fluid distribution in RR gravity.

In the limit of $\Xi=\Theta=1$, it is worth noticing that Eq.~\eqref{eq:YTF} is slightly different from Eq.~(49) in Ref.~\cite{Nazar:2021hgg}, where the authors focused on the complexity factor in Rastall gravity. The discrepancy arises from an error during their calculation of the Weyl tensor.

\section{Stellar models with vanishing complexity factor}\label{sec:vancom}

Historically, in order to provide theoretical explanations for astrophysical objects, it was commonly assumed that matter was composed of isotropic perfect fluids. However, realistic stellar interior may possess anisotropic matter distributions \cite{Lemaitre:1933gd}, where the radial pressure $p_{r}^{}$ is not equal to the tangential pressure $p_{t}^{}$. Especially for high-density configurations, Ruderman pointed out that pressure anisotropy could arise from nuclear matter interactions \cite{Ruderman:1972aj}. Hillebrandt and Steinmetz also showed that anisotropy cannot be neglected when modeling neutron stars \cite{Hillebrandt1976}. It has been shown in \cite{Bowers:1974tgi} that the presence of anisotropic pressure is more conducive to generate objects with high redshift. In fact, anisotropy can be induced by various physical processes, such as phase transitions \cite{Sokolov1980}, pion condensates \cite{Sawyer:1972cq} or other phenomena. Recently, anisotropic fluid is used in the study of the structure of fuzzy dark matter \cite{Khan:2024vsh}. We refer readers to Refs.~\cite{Herrera:1997plx,Mak:2001eb} for a comprehensive survey on this topic.

To solve the stellar structure and spacetime geometry for a static, spherically symmetric and anisotropic fluid distribution in RR gravity, there are five unknown functions, $\left(\alpha,\beta,\rho,p_{r}^{},p_{t}^{}\right)$, that need to be determined. The theory itself, however, only provides three independent differential equations, i.e., Eqs.~\eqref{eq:zerozero}, \eqref{eq:oneone} and \eqref{eq:twotwo}. One has to supplement two equations to close the system of equations. Actually, after the complexity factor is defined, it is natural to set $Y_{\text{TF}}^{}=0$ as one of them \cite{Herrera:2018bww}. From Eq.~\eqref{eq:YTF}, the vanishing complexity factor condition is
\begin{equation}\label{eq:VCC}
8\pi\Pi=\frac{1}{2r^{3}}\int_{0}^{r}x^{3}\left(8\pi\rho-\frac{1-k}{2}R\right)^{\prime}dx,
\end{equation}
which might be regarded as a non-local equation of state. The complexity factor and gravitational decoupling technique \cite{Ovalle:2017fgl} are often integrated for constructing uncharged or charged analytical solutions in astrophysics. For recent advances, see, e.g., Refs.~\cite{Albalahi:2024vpy,Albalahi:2024ujg,Yousaf:2024fkr}.

For convenience in application, we are supposed to convert it into a metric constraint. The left-hand side can be calculated by subtracting \eqref{eq:oneone} from \eqref{eq:twotwo}, while an integral by parts with 00-component \eqref{eq:zerozero} gives
\begin{equation}\label{eq:rhs}
\text{r.h.s of Eq.~\eqref{eq:VCC}}=\frac{\Theta^{2}}{e^{2\beta}}\frac{1-e^{2\beta}+r\beta^{\prime}}{r^{2}}.
\end{equation}
Hence, the vanishing complexity condition is equivalent to the Ricatti equation for $\alpha^{\prime}$ as
\begin{equation}\label{eq:VCC0}
\alpha^{\prime}=r\left(\alpha^{\prime\prime}-\alpha^{\prime}\beta^{\prime}+\alpha^{\prime 2}\right),
\end{equation}
and the solution is
\begin{equation}\label{eq:VCC1}
e^{\alpha}=A_{1}+A_{2}\int_{0}^{r}xe^{\beta}dx,
\end{equation}
or reversely
\begin{equation}
e^{\beta}=\frac{B}{r}\frac{d}{dr}e^{\alpha},
\end{equation}
where $A_{1}$, $A_{2}$ and $B$ are integral constants. Thus, the vanishing complexity condition of RR gravity is identical to that of GR \cite{Contreras:2022vec}.

Now, we still need another equation, which is usually either specific form of metric functions, or equation of state. For instance, the Finch-Skea potential \cite{Finch_1989}
\begin{equation}\label{eq:FS}
e^{2\beta}=1+Kr^{2}
\end{equation}
with $K$ is a constant of dimension $\operatorname{L}^{-2}$. Eq.~\eqref{eq:VCC1} under this ansatz leads to
\begin{equation}
e^{\alpha}=A_{1}+\frac{A_{2}}{3K}\left(1+Kr^{2}\right)^{3/2}.
\end{equation}
Next, by solving $\rho$, $p_{r}^{}$ and $p_{t}^{}$ from Eqs.~\eqref{eq:zerozero}, \eqref{eq:oneone} and \eqref{eq:twotwo}, respectively, and using the boundary conditions \eqref{eq:BC} to fix constants $A_{1}$, $A_{2}$ and $K$, the whole system is eventually completely determined. A comprehensive analysis for this kind of solutions exceeds from the scopes of the current work, the reader is invited to consult Refs.~\cite{Habsi:2023stx,Andrade:2023jvm,Khan:2024jnb} for the applications of Eq.~\eqref{eq:FS} in the complexity framework.

It is interesting to find that, from Eq.\eqref{eq:VCC}, in the isotropic situation, i.e. $\Pi =0$, there may exist non-uniform stellar solutions, which is radically different from those in GR.  

Let $\Pi=0$, Eqs.~\eqref{eq:VCC} and \eqref{eq:rhs} imply that
\begin{equation}
e^{-2\beta}=1-H^{2}r^{2},
\end{equation}
where $H$ is an integral constant. Then Eq.~\eqref{eq:VCC1} yields
\begin{equation}
e^{2\alpha}=\left(A_{1}-A_{2}H^{-2}\sqrt{1-H^{2}r^{2}}\right)^{2}.
\end{equation}
Substituting these results back into Eqs.~\eqref{eq:zerozero} and \eqref{eq:oneone}, we have
\begin{align}
\rho={}&\frac{3\Theta^{2}}{8\pi}\left(A_{1}-A_{2}H^{-2}\sqrt{1-H^{2}r^{2}}\right)^{-2}\notag\\
&\left[A_{1}^{2}H^{2}k+A_{1}A_{2}\left(1-3k\right)\sqrt{1-H^{2}r^{2}}\right.\notag\\
&\left.\quad-\,A_{2}^{2}H^{-2}\left(1-2k\right)\left(1-H^{2}r^{2}\right)\phantom{\hspace{-2.3em}\sqrt{H^{2}}}\right],\\
p={}&\frac{\Theta^{2}}{8\pi}\left(A_{1}-A_{2}H^{-2}\sqrt{1-H^{2}r^{2}}\right)^{-2}\notag\\
&\left[A_{1}^{2}H^{2}\left(2-3k\right)-A_{1}A_{2}\left(5-9k\right)\sqrt{1-H^{2}r^{2}}\right.\notag\\
&\left.\quad+\,3A_{2}^{2}H^{-2}\left(1-2k\right)\left(1-H^{2}r^{2}\right)\phantom{\hspace{-2.3em}\sqrt{H^{2}}}\right],\label{eq:p}
\end{align}
where $p\equiv p_{r}^{}=p_{t}^{}$. From Eq.\eqref{eq:BC}, the constants are found to be
\begin{gather}
H^{2}=\frac{2M}{R_{\Sigma}^{3}},\\
A_{1}=\frac{3\left(1-2k\right)\sqrt{1-H^{2}R_{\Sigma}^{2}}}{1-3k},\\
A_{2}=\frac{\left(2-3k\right)H^{2}}{1-3k}.
\end{gather}

This configuration is supported by five parameters, and reduces to the Schwarzschild interior solution \cite{Schwarzschild:1916ae} when we take $k=\Theta=\Xi=1$. 

\begin{figure}[htbp]
\includegraphics[width=0.475\textwidth]{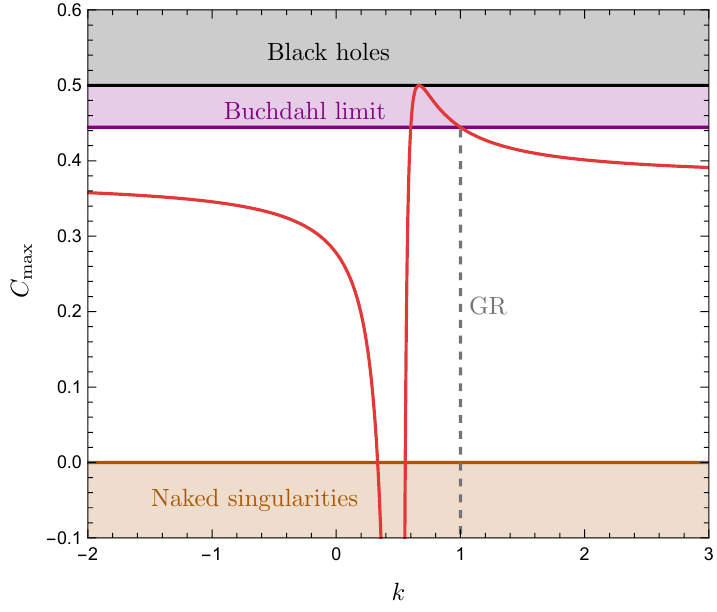}
\caption{Maximum compactness versus dimensionless Rastall parameter. The well-known Buchdahl bound $C_{\text{max}}<4/9$ \cite{Buchdahl:1959zz} in GR is reproduced by setting $k=1$.}
\label{fig:Cmaxofk}
\end{figure}

To enforce regularity of the model at $r=0$, from Eq.~\eqref{eq:p}  the compactness of the object $C\equiv M/R_{\Sigma}^{}$ should be less than 
\begin{equation}
C_{\text{max}} = \frac{\left(1-3k\right)\left(5-9k\right)}{18\left(1-2k\right)^{2}}.
\end{equation}
Therefore, the dimensionless Rastall parameters $k \in [1/3,5/9]$ are ruled out by the vanishing complexity condition.
Interestingly, for a given value of Rastall parameter $k \in ({3}/{5}, 1)$, an ultra-compact isotropic object beyond Buchdahl bound can be obtained. Furthermore, when $k = {2}/{3}$, $C_{\text{max}}$ takes its maximal value ${1}/{2}$, which is just the value of compactness for a static black hole  as shown in Figure~\ref{fig:Cmaxofk}. 

\section{Conclusions}\label{sec:conclu}
In the present work, we generalized the definition of complexity factor for static spherically symmetric fluid distributions from GR to RR theory of gravity. As in the case of GR \cite{Herrera:2018bww}, the complexity factor $Y_{\text{TF}}^{}$ is also one of the structure scalars, which appear in the orthogonal splitting of the Riemann curvature tensor. It is found that $Y_{\text{TF}}^{}$ characterizes how the value of active gravitational mass is affected by the inhomogeneous energy density and anisotropic pressure of the fluid distribution in the context of RR gravity.

When the two rainbow functions $\Xi$ and $\Theta$ and Rastall's $k$ parameter are both taken to be one, all of our results clearly reduce to Herrera \cite{Herrera:2018bww}, as expected. Moreover, in the low-energy limit, our work corrects the computation for Rastall theory alone performed in Ref.~\cite{Nazar:2021hgg}, which is somewhat flawed.

Additionally, we recast the vanishing complexity factor condition into a much more user-friendly form, and a specific metric potential is employed to illustrate how to build an anisotropic stellar model within this framework, for the sake of completeness. Also, we provide an isotropic solution featuring minimal complexity, which is a variable-density extension of the Schwarzschild stars. For a small region of the parameter space, the radius of the model can be arbitrarily close to event horizon $2M$ without developing any singularities in spacetime, and it  may serve as a black-hole mimicker. Further investigations on other aspects, such as the physical acceptability, are under preparation.

The expression of the complexity factor $Y_{\text{TF}}^{}$ in RR gravity is different from one in GR indeed, however, a glance at Eq.~\eqref{eq:VCC} or \eqref{eq:VCC0} confirms that the vanishing complexity condition $Y_{\text{TF}}^{}=0$ is theory-independent due to the fact that the term  $8\pi\rho-\left(1-k\right)R/2$ plays the role of effective energy density. As a result, the bridge equation \eqref{eq:VCC1} linking two metric functions can be easily used to explore the stellar structures in RR gravity.
\section*{Acknowledgement}
The authors are indebted to Fei Xie for their careful check of the calculations, and to Yu-Bo Peng for providing us with some literature. This work is supported by the innovation program of Shanghai Normal University under Grant No.~KF202147.
\bibliography{ref}

\begin{thebibliography}{85}
\expandafter\ifx\csname natexlab\endcsname\relax\def\natexlab#1{#1}\fi
\providecommand{\url}[1]{\texttt{#1}}
\providecommand{\href}[2]{#2}
\providecommand{\path}[1]{#1}
\providecommand{\DOIprefix}{doi:}
\providecommand{\ArXivprefix}{arXiv:}
\providecommand{\URLprefix}{URL: }
\providecommand{\Pubmedprefix}{pmid:}
\providecommand{\doi}[1]{\href{http://dx.doi.org/#1}{\path{#1}}}
\providecommand{\Pubmed}[1]{\href{pmid:#1}{\path{#1}}}
\providecommand{\bibinfo}[2]{#2}
\ifx\xfnm\relax \def\xfnm[#1]{\unskip,\space#1}\fi
\bibitem[{Kolmogorov(1968)}]{Kolmogorov}
\bibinfo{author}{A.~N. Kolmogorov},
\newblock \bibinfo{title}{Three approaches to the quantitative definition of
  information},
\newblock \bibinfo{journal}{Int J Comput Math} \bibinfo{volume}{2}
  (\bibinfo{year}{1968}) \bibinfo{pages}{157--168}.
  \DOIprefix\doi{10.1080/00207166808803030}.
\bibitem[{Grassberger(1986)}]{Grassberger1986}
\bibinfo{author}{P.~Grassberger},
\newblock \bibinfo{title}{Toward a quantitative theory of self-generated
  complexity},
\newblock \bibinfo{journal}{Int. J. Theor. Phys.} \bibinfo{volume}{25}
  (\bibinfo{year}{1986}) \bibinfo{pages}{907--938}.
  \DOIprefix\doi{10.1007/BF00668821}.
\bibitem[{Lloyd and Pagels(1988)}]{Lloyd:1988cn}
\bibinfo{author}{S.~Lloyd}, \bibinfo{author}{H.~Pagels},
\newblock \bibinfo{title}{{Complexity as thermodynamic depth}},
\newblock \bibinfo{journal}{Annals Phys.} \bibinfo{volume}{188}
  (\bibinfo{year}{1988}) \bibinfo{pages}{186}.
  \DOIprefix\doi{10.1016/0003-4916(88)90094-2}.
\bibitem[{Crutchfield and Young(1989)}]{PhysRevLett.63.105}
\bibinfo{author}{J.~P. Crutchfield}, \bibinfo{author}{K.~Young},
\newblock \bibinfo{title}{Inferring statistical complexity},
\newblock \bibinfo{journal}{Phys. Rev. Lett.} \bibinfo{volume}{63}
  (\bibinfo{year}{1989}) \bibinfo{pages}{105--108}.
  \DOIprefix\doi{10.1103/PhysRevLett.63.105}.
\bibitem[{Anderson(1991)}]{10.1063/1.2810163}
\bibinfo{author}{P.~W. Anderson},
\newblock \bibinfo{title}{{Is Complexity Physics? Is It Science? What is It?}},
\newblock \bibinfo{journal}{Phys. Today} \bibinfo{volume}{44}
  (\bibinfo{year}{1991}) \bibinfo{pages}{9--11}.
  \DOIprefix\doi{10.1063/1.2810163}.
\bibitem[{Parisi(1993)}]{Parisi_1993}
\bibinfo{author}{G.~Parisi},
\newblock \bibinfo{title}{Statistical physics and biology},
\newblock \bibinfo{journal}{Phys. World} \bibinfo{volume}{6}
  (\bibinfo{year}{1993}) \bibinfo{pages}{42}.
  \DOIprefix\doi{10.1088/2058-7058/6/9/35}.
\bibitem[{López-Ruiz et~al.(1995)López-Ruiz, Mancini, and
  Calbet}]{LOPEZRUIZ1995321}
\bibinfo{author}{R.~López-Ruiz}, \bibinfo{author}{H.~Mancini},
  \bibinfo{author}{X.~Calbet},
\newblock \bibinfo{title}{A statistical measure of complexity},
\newblock \bibinfo{journal}{Phys. Lett. A} \bibinfo{volume}{209}
  (\bibinfo{year}{1995}) \bibinfo{pages}{321--326}.
  \DOIprefix\doi{10.1016/0375-9601(95)00867-5}.
\bibitem[{Feldman and Crutchfield(1998)}]{FELDMAN1998244}
\bibinfo{author}{D.~P. Feldman}, \bibinfo{author}{J.~P. Crutchfield},
\newblock \bibinfo{title}{Measures of statistical complexity: Why?},
\newblock \bibinfo{journal}{Phys. Lett. A} \bibinfo{volume}{238}
  (\bibinfo{year}{1998}) \bibinfo{pages}{244--252}.
  \DOIprefix\doi{10.1016/S0375-9601(97)00855-4}.
\bibitem[{Calbet and L\'opez-Ruiz(2001)}]{PhysRevE.63.066116}
\bibinfo{author}{X.~Calbet}, \bibinfo{author}{R.~L\'opez-Ruiz},
\newblock \bibinfo{title}{Tendency towards maximum complexity in a
  nonequilibrium isolated system},
\newblock \bibinfo{journal}{Phys. Rev. E} \bibinfo{volume}{63}
  (\bibinfo{year}{2001}) \bibinfo{pages}{066116}.
  \DOIprefix\doi{10.1103/PhysRevE.63.066116}.
\bibitem[{Catal\'an et~al.(2002)Catal\'an, Garay, and
  L\'opez-Ruiz}]{PhysRevE.66.011102}
\bibinfo{author}{R.~G. Catal\'an}, \bibinfo{author}{J.~Garay},
  \bibinfo{author}{R.~L\'opez-Ruiz},
\newblock \bibinfo{title}{Features of the extension of a statistical measure of
  complexity to continuous systems},
\newblock \bibinfo{journal}{Phys. Rev. E} \bibinfo{volume}{66}
  (\bibinfo{year}{2002}) \bibinfo{pages}{011102}.
  \DOIprefix\doi{10.1103/PhysRevE.66.011102}.
\bibitem[{Sañudo and López-Ruiz(2008)}]{SANUDO20085283}
\bibinfo{author}{J.~Sañudo}, \bibinfo{author}{R.~López-Ruiz},
\newblock \bibinfo{title}{Statistical complexity and fisher-shannon information
  in the h-atom},
\newblock \bibinfo{journal}{Phys. Lett. A} \bibinfo{volume}{372}
  (\bibinfo{year}{2008}) \bibinfo{pages}{5283--5286}.
  \DOIprefix\doi{10.1016/j.physleta.2008.06.012}.
\bibitem[{Panos et~al.(2009)Panos, Nikolaidis, Chatzisavvas, and
  Tsouros}]{PANOS20092343}
\bibinfo{author}{C.~Panos}, \bibinfo{author}{N.~Nikolaidis},
  \bibinfo{author}{K.~Chatzisavvas}, \bibinfo{author}{C.~Tsouros},
\newblock \bibinfo{title}{A simple method for the evaluation of the information
  content and complexity in atoms. a proposal for scalability},
\newblock \bibinfo{journal}{Phys. Lett. A} \bibinfo{volume}{373}
  (\bibinfo{year}{2009}) \bibinfo{pages}{2343--2350}.
  \DOIprefix\doi{10.1016/j.physleta.2009.04.070}.
\bibitem[{Susskind(2016)}]{Susskind:2014moa}
\bibinfo{author}{L.~Susskind},
\newblock \bibinfo{title}{{Entanglement is not enough}},
\newblock \bibinfo{journal}{Fortsch. Phys.} \bibinfo{volume}{64}
  (\bibinfo{year}{2016}) \bibinfo{pages}{49--71}.
  \DOIprefix\doi{10.1002/prop.201500095}.
\bibitem[{Brown et~al.(2016)Brown, Roberts, Susskind, Swingle, and
  Zhao}]{Brown:2015lvg}
\bibinfo{author}{A.~R. Brown}, \bibinfo{author}{D.~A. Roberts},
  \bibinfo{author}{L.~Susskind}, \bibinfo{author}{B.~Swingle},
  \bibinfo{author}{Y.~Zhao},
\newblock \bibinfo{title}{{Complexity, action, and black holes}},
\newblock \bibinfo{journal}{Phys. Rev. D} \bibinfo{volume}{93}
  (\bibinfo{year}{2016}) \bibinfo{pages}{086006}.
  \DOIprefix\doi{10.1103/PhysRevD.93.086006}.
\bibitem[{Sanudo and Pacheco(2009)}]{Sanudo:2008bu}
\bibinfo{author}{J.~Sanudo}, \bibinfo{author}{A.~F. Pacheco},
\newblock \bibinfo{title}{{Complexity and white-dwarf structure}},
\newblock \bibinfo{journal}{Phys. Lett. A} \bibinfo{volume}{373}
  (\bibinfo{year}{2009}) \bibinfo{pages}{807--810}.
  \DOIprefix\doi{10.1016/j.physleta.2009.01.008}.
\bibitem[{Chatzisavvas et~al.(2009)Chatzisavvas, Psonis, Panos, and
  Moustakidis}]{Chatzisavvas:2009xu}
\bibinfo{author}{K.~C. Chatzisavvas}, \bibinfo{author}{V.~P. Psonis},
  \bibinfo{author}{C.~P. Panos}, \bibinfo{author}{C.~C. Moustakidis},
\newblock \bibinfo{title}{{Complexity and neutron stars structure}},
\newblock \bibinfo{journal}{Phys. Lett. A} \bibinfo{volume}{373}
  (\bibinfo{year}{2009}) \bibinfo{pages}{3901--3909}.
  \DOIprefix\doi{10.1016/j.physleta.2009.08.042}.
\bibitem[{De~Avellar and Horvath(2012)}]{DeAvellar:2012zz}
\bibinfo{author}{M.~G.~B. De~Avellar}, \bibinfo{author}{J.~E. Horvath},
\newblock \bibinfo{title}{{Entropy, complexity and disequilibrium in compact
  stars}},
\newblock \bibinfo{journal}{Phys. Lett. A} \bibinfo{volume}{376}
  (\bibinfo{year}{2012}) \bibinfo{pages}{1085--1089}.
  \DOIprefix\doi{10.1016/j.physleta.2012.02.012}.
\bibitem[{Herrera(2018)}]{Herrera:2018bww}
\bibinfo{author}{L.~Herrera},
\newblock \bibinfo{title}{{New definition of complexity for self-gravitating
  fluid distributions: The spherically symmetric, static case}},
\newblock \bibinfo{journal}{Phys. Rev. D} \bibinfo{volume}{97}
  (\bibinfo{year}{2018}) \bibinfo{pages}{044010}.
  \DOIprefix\doi{10.1103/PhysRevD.97.044010}.
\bibitem[{Sharif and Butt(2018)}]{Sharif:2018pgq}
\bibinfo{author}{M.~Sharif}, \bibinfo{author}{I.~I. Butt},
\newblock \bibinfo{title}{{Complexity Factor for Charged Spherical System}},
\newblock \bibinfo{journal}{Eur. Phys. J. C} \bibinfo{volume}{78}
  (\bibinfo{year}{2018}) \bibinfo{pages}{688}.
  \DOIprefix\doi{10.1140/epjc/s10052-018-6121-5}.
\bibitem[{Herrera et~al.(2018)Herrera, Di~Prisco, and Ospino}]{Herrera:2018czt}
\bibinfo{author}{L.~Herrera}, \bibinfo{author}{A.~Di~Prisco},
  \bibinfo{author}{J.~Ospino},
\newblock \bibinfo{title}{{Definition of complexity for dynamical spherically
  symmetric dissipative self-gravitating fluid distributions}},
\newblock \bibinfo{journal}{Phys. Rev. D} \bibinfo{volume}{98}
  (\bibinfo{year}{2018}) \bibinfo{pages}{104059}.
  \DOIprefix\doi{10.1103/PhysRevD.98.104059}.
\bibitem[{Sharif and Butt(2018)}]{Sharif:2018efi}
\bibinfo{author}{M.~Sharif}, \bibinfo{author}{I.~I. Butt},
\newblock \bibinfo{title}{{Complexity factor for static cylindrical system}},
\newblock \bibinfo{journal}{Eur. Phys. J. C} \bibinfo{volume}{78}
  (\bibinfo{year}{2018}) \bibinfo{pages}{850}.
  \DOIprefix\doi{10.1140/epjc/s10052-018-6330-y}.
\bibitem[{Herrera et~al.(2019)Herrera, Di~Prisco, and Ospino}]{Herrera:2019cbx}
\bibinfo{author}{L.~Herrera}, \bibinfo{author}{A.~Di~Prisco},
  \bibinfo{author}{J.~Ospino},
\newblock \bibinfo{title}{{Complexity factors for axially symmetric static
  sources}},
\newblock \bibinfo{journal}{Phys. Rev. D} \bibinfo{volume}{99}
  (\bibinfo{year}{2019}) \bibinfo{pages}{044049}.
  \DOIprefix\doi{10.1103/PhysRevD.99.044049}.
\bibitem[{Abbas and Nazar(2018)}]{Abbas:2018idr}
\bibinfo{author}{G.~Abbas}, \bibinfo{author}{H.~Nazar},
\newblock \bibinfo{title}{{Complexity Factor For Static Anisotropic
  Self-Gravitating Source in $f(R)$ Gravity}},
\newblock \bibinfo{journal}{Eur. Phys. J. C} \bibinfo{volume}{78}
  (\bibinfo{year}{2018}) \bibinfo{pages}{510}.
  \DOIprefix\doi{10.1140/epjc/s10052-018-5973-z}.
\bibitem[{Sharif and Majid(2019)}]{Sharif:2019efn}
\bibinfo{author}{M.~Sharif}, \bibinfo{author}{A.~Majid},
\newblock \bibinfo{title}{{Complexity factor for static sphere in
  self-interacting Brans\textendash{}Dicke gravity}},
\newblock \bibinfo{journal}{Chin. J. Phys.} \bibinfo{volume}{61}
  (\bibinfo{year}{2019}) \bibinfo{pages}{38--46}.
  \DOIprefix\doi{10.1016/j.cjph.2019.08.004}.
\bibitem[{Sharif et~al.(2019)Sharif, Majid, and Nasir}]{Sharif:2019sze}
\bibinfo{author}{M.~Sharif}, \bibinfo{author}{A.~Majid},
  \bibinfo{author}{M.~M.~M. Nasir},
\newblock \bibinfo{title}{{Complexity factor for self-gravitating system in
  modified Gauss\textendash{}Bonnet gravity}},
\newblock \bibinfo{journal}{Int. J. Mod. Phys. A} \bibinfo{volume}{34}
  (\bibinfo{year}{2019}) \bibinfo{pages}{1950210}.
  \DOIprefix\doi{10.1142/S0217751X19502105}.
\bibitem[{Yousaf et~al.(2020{\natexlab{a}})Yousaf, Bhatti, and
  Naseer}]{Yousaf:2020xux}
\bibinfo{author}{Z.~Yousaf}, \bibinfo{author}{M.~Z. Bhatti},
  \bibinfo{author}{T.~Naseer},
\newblock \bibinfo{title}{{New definition of complexity factor in
  $f(R,T,R_{\mu\nu}T^{\mu\nu})$ gravity}},
\newblock \bibinfo{journal}{Phys. Dark Univ.} \bibinfo{volume}{28}
  (\bibinfo{year}{2020}{\natexlab{a}}) \bibinfo{pages}{100535}.
  \DOIprefix\doi{10.1016/j.dark.2020.100535}.
\bibitem[{Yousaf et~al.(2020{\natexlab{b}})Yousaf, Khlopov, Bhatti, and
  Naseer}]{Yousaf:2020xft}
\bibinfo{author}{Z.~Yousaf}, \bibinfo{author}{M.~Y. Khlopov},
  \bibinfo{author}{M.~Z. Bhatti}, \bibinfo{author}{T.~Naseer},
\newblock \bibinfo{title}{{Influence of Modification of Gravity on the
  Complexity Factor of Static Spherical Structures}},
\newblock \bibinfo{journal}{Mon. Not. Roy. Astron. Soc.} \bibinfo{volume}{495}
  (\bibinfo{year}{2020}{\natexlab{b}}) \bibinfo{pages}{4334--4346}.
  \DOIprefix\doi{10.1093/mnras/staa1470}.
\bibitem[{Nazar et~al.(2021)Nazar, Alkhaldi, Abbas, and
  Shahzad}]{Nazar:2021hgg}
\bibinfo{author}{H.~Nazar}, \bibinfo{author}{A.~H. Alkhaldi},
  \bibinfo{author}{G.~Abbas}, \bibinfo{author}{M.~R. Shahzad},
\newblock \bibinfo{title}{{Complexity factor for anisotropic self-gravitating
  sphere in Rastall gravity}},
\newblock \bibinfo{journal}{Int. J. Mod. Phys. A} \bibinfo{volume}{36}
  (\bibinfo{year}{2021}) \bibinfo{pages}{2150233}.
  \DOIprefix\doi{10.1142/S0217751X2150233X}.
\bibitem[{Bhatti et~al.(2021{\natexlab{a}})Bhatti, Khlopov, Yousaf, and
  Khan}]{Bhatti:2021jue}
\bibinfo{author}{M.~Z. Bhatti}, \bibinfo{author}{M.~Y. Khlopov},
  \bibinfo{author}{Z.~Yousaf}, \bibinfo{author}{S.~Khan},
\newblock \bibinfo{title}{{Electromagnetic field and complexity of relativistic
  fluids in f (G) gravity}},
\newblock \bibinfo{journal}{Mon. Not. Roy. Astron. Soc.} \bibinfo{volume}{506}
  (\bibinfo{year}{2021}{\natexlab{a}}) \bibinfo{pages}{4543--4560}.
  \DOIprefix\doi{10.1093/mnras/stab2062}.
\bibitem[{Bhatti et~al.(2021{\natexlab{b}})Bhatti, Yousaf, and
  Khan}]{Bhatti:2021pxr}
\bibinfo{author}{M.~Z. Bhatti}, \bibinfo{author}{Z.~Yousaf},
  \bibinfo{author}{S.~Khan},
\newblock \bibinfo{title}{{Role of quasi-homologous condition to study complex
  systems in $f({\mathbb {G}}, T)$ gravity}},
\newblock \bibinfo{journal}{Eur. Phys. J. Plus} \bibinfo{volume}{136}
  (\bibinfo{year}{2021}{\natexlab{b}}) \bibinfo{pages}{975}.
  \DOIprefix\doi{10.1140/epjp/s13360-021-01889-9}.
\bibitem[{Yousaf et~al.(2022)Yousaf, Bhatti, Khan, and Sahoo}]{Yousaf:2021xex}
\bibinfo{author}{Z.~Yousaf}, \bibinfo{author}{M.~Z. Bhatti},
  \bibinfo{author}{S.~Khan}, \bibinfo{author}{P.~K. Sahoo},
\newblock
  \bibinfo{title}{{f(G,T\ensuremath{\alpha}\ensuremath{\beta}T\ensuremath{\alpha}\ensuremath{\beta})
  theory and complex cosmological structures}},
\newblock \bibinfo{journal}{Phys. Dark Univ.} \bibinfo{volume}{36}
  (\bibinfo{year}{2022}) \bibinfo{pages}{101015}.
  \DOIprefix\doi{10.1016/j.dark.2022.101015}.
\bibitem[{Sharif and Ahmad(2022)}]{Sharif:2022tbp}
\bibinfo{author}{M.~Sharif}, \bibinfo{author}{K.~Ahmad},
\newblock \bibinfo{title}{{Complexity of static sphere in
  energy\textendash{}momentum squared gravity}},
\newblock \bibinfo{journal}{Mod. Phys. Lett. A} \bibinfo{volume}{37}
  (\bibinfo{year}{2022}) \bibinfo{pages}{2250031}.
  \DOIprefix\doi{10.1142/S0217732322500316}.
\bibitem[{Bhatti et~al.(2022)Bhatti, Yousaf, and Hanif}]{Bhatti:2022pqe}
\bibinfo{author}{M.~Z. Bhatti}, \bibinfo{author}{Z.~Yousaf},
  \bibinfo{author}{S.~Hanif},
\newblock \bibinfo{title}{{A novel definition of complexity in torsion based
  theory}},
\newblock \bibinfo{journal}{Eur. Phys. J. C} \bibinfo{volume}{82}
  (\bibinfo{year}{2022}) \bibinfo{pages}{714}.
  \DOIprefix\doi{10.1140/epjc/s10052-022-10688-0}.
\bibitem[{Rastall(1972)}]{Rastall:1972swe}
\bibinfo{author}{P.~Rastall},
\newblock \bibinfo{title}{{Generalization of the einstein theory}},
\newblock \bibinfo{journal}{Phys. Rev. D} \bibinfo{volume}{6}
  (\bibinfo{year}{1972}) \bibinfo{pages}{3357--3359}.
  \DOIprefix\doi{10.1103/PhysRevD.6.3357}.
\bibitem[{Rastall(1976)}]{Rastall:1976uh}
\bibinfo{author}{P.~Rastall},
\newblock \bibinfo{title}{{A Theory of Gravity}},
\newblock \bibinfo{journal}{Can. J. Phys.} \bibinfo{volume}{54}
  (\bibinfo{year}{1976}) \bibinfo{pages}{66--75}.
  \DOIprefix\doi{10.1139/p76-008}.
\bibitem[{Visser(2018)}]{Visser:2017gpz}
\bibinfo{author}{M.~Visser},
\newblock \bibinfo{title}{{Rastall gravity is equivalent to Einstein gravity}},
\newblock \bibinfo{journal}{Phys. Lett. B} \bibinfo{volume}{782}
  (\bibinfo{year}{2018}) \bibinfo{pages}{83--86}.
  \DOIprefix\doi{10.1016/j.physletb.2018.05.028}.
\bibitem[{Darabi et~al.(2018)Darabi, Moradpour, Licata, Heydarzade, and
  Corda}]{Darabi:2017coc}
\bibinfo{author}{F.~Darabi}, \bibinfo{author}{H.~Moradpour},
  \bibinfo{author}{I.~Licata}, \bibinfo{author}{Y.~Heydarzade},
  \bibinfo{author}{C.~Corda},
\newblock \bibinfo{title}{{Einstein and Rastall Theories of Gravitation in
  Comparison}},
\newblock \bibinfo{journal}{Eur. Phys. J. C} \bibinfo{volume}{78}
  (\bibinfo{year}{2018}) \bibinfo{pages}{25}.
  \DOIprefix\doi{10.1140/epjc/s10052-017-5502-5}.
\bibitem[{Smalley(1983)}]{Smalley_1983}
\bibinfo{author}{L.~L. Smalley},
\newblock \bibinfo{title}{Rastall's and related theories are conservative
  gravitational theories although physically inequivalent to general
  relativity},
\newblock \bibinfo{journal}{J. Phys. A: Math. Gen.} \bibinfo{volume}{16}
  (\bibinfo{year}{1983}) \bibinfo{pages}{2179}.
  \DOIprefix\doi{10.1088/0305-4470/16/10/014}.
\bibitem[{Al-Rawaf and Taha(1996)}]{Al-Rawaf:1995xkt}
\bibinfo{author}{A.~S. Al-Rawaf}, \bibinfo{author}{M.~O. Taha},
\newblock \bibinfo{title}{{A Resolution of the cosmological age puzzle}},
\newblock \bibinfo{journal}{Phys. Lett. B} \bibinfo{volume}{366}
  (\bibinfo{year}{1996}) \bibinfo{pages}{69--71}.
  \DOIprefix\doi{10.1016/0370-2693(95)01145-5}.
\bibitem[{Maulana and Sulaksono(2019)}]{Maulana:2019evy}
\bibinfo{author}{H.~Maulana}, \bibinfo{author}{A.~Sulaksono},
\newblock \bibinfo{title}{{Energy condition in Rastall gravity theory}},
\newblock \bibinfo{journal}{J. Phys. Conf. Ser.} \bibinfo{volume}{1321}
  (\bibinfo{year}{2019}) \bibinfo{pages}{022012}.
  \DOIprefix\doi{10.1088/1742-6596/1321/2/022012}.
\bibitem[{Li et~al.(2019)Li, Wang, Xu, and Guo}]{Li:2019jkv}
\bibinfo{author}{R.~Li}, \bibinfo{author}{J.~Wang}, \bibinfo{author}{Z.~Xu},
  \bibinfo{author}{X.~Guo},
\newblock \bibinfo{title}{{Constraining the Rastall parameters in static
  space\textendash{}times with galaxy-scale strong gravitational lensing}},
\newblock \bibinfo{journal}{Mon. Not. Roy. Astron. Soc.} \bibinfo{volume}{486}
  (\bibinfo{year}{2019}) \bibinfo{pages}{2407--2411}.
  \DOIprefix\doi{10.1093/mnras/stz967}.
\bibitem[{Manna et~al.(2019)Manna, Rahaman, and Mondal}]{Manna:2019onx}
\bibinfo{author}{T.~Manna}, \bibinfo{author}{F.~Rahaman},
  \bibinfo{author}{M.~Mondal},
\newblock \bibinfo{title}{{Solar system tests in Rastall gravity}},
\newblock \bibinfo{journal}{Mod. Phys. Lett. A} \bibinfo{volume}{35}
  (\bibinfo{year}{2019}) \bibinfo{pages}{2050034}.
  \DOIprefix\doi{10.1142/S0217732320500340}.
\bibitem[{Saleem and Hassan(2020)}]{Saleem:2020bwa}
\bibinfo{author}{R.~Saleem}, \bibinfo{author}{J.~Hassan},
\newblock \bibinfo{title}{{Confronting the warm vector inflation in Rastall
  theory of gravity with Planck 2018 data}},
\newblock \bibinfo{journal}{Phys. Dark Univ.} \bibinfo{volume}{28}
  (\bibinfo{year}{2020}) \bibinfo{pages}{100515}.
  \DOIprefix\doi{10.1016/j.dark.2020.100515}.
\bibitem[{El~Hanafy(2022)}]{ElHanafy:2022kjl}
\bibinfo{author}{W.~El~Hanafy},
\newblock \bibinfo{title}{{Impact of Rastall Gravity on Mass, Radius, and Sound
  Speed of the Pulsar PSR J0740+6620}},
\newblock \bibinfo{journal}{Astrophys. J.} \bibinfo{volume}{940}
  (\bibinfo{year}{2022}) \bibinfo{pages}{51}.
  \DOIprefix\doi{10.3847/1538-4357/ac9410}.
\bibitem[{Saleem and Saleem(2024)}]{Saleem:2024ojw}
\bibinfo{author}{R.~Saleem}, \bibinfo{author}{A.~Saleem},
\newblock \bibinfo{title}{{Constraining Rastall parameter for baryon asymmetry
  factor using isotropic and anisotropic universe models}},
\newblock \bibinfo{journal}{Int. J. Geom. Meth. Mod. Phys.}
  \bibinfo{volume}{21} (\bibinfo{year}{2024}) \bibinfo{pages}{2450102}.
  \DOIprefix\doi{10.1142/S0219887824501020}.
\bibitem[{Magueijo and Smolin(2004)}]{Magueijo:2002xx}
\bibinfo{author}{J.~Magueijo}, \bibinfo{author}{L.~Smolin},
\newblock \bibinfo{title}{{Gravity's rainbow}},
\newblock \bibinfo{journal}{Class. Quant. Grav.} \bibinfo{volume}{21}
  (\bibinfo{year}{2004}) \bibinfo{pages}{1725--1736}.
  \DOIprefix\doi{10.1088/0264-9381/21/7/001}.
\bibitem[{Bruno et~al.(2001)Bruno, Amelino-Camelia, and
  Kowalski-Glikman}]{Bruno:2001mw}
\bibinfo{author}{N.~R. Bruno}, \bibinfo{author}{G.~Amelino-Camelia},
  \bibinfo{author}{J.~Kowalski-Glikman},
\newblock \bibinfo{title}{{Deformed boost transformations that saturate at the
  Planck scale}},
\newblock \bibinfo{journal}{Phys. Lett. B} \bibinfo{volume}{522}
  (\bibinfo{year}{2001}) \bibinfo{pages}{133--138}.
  \DOIprefix\doi{10.1016/S0370-2693(01)01264-3}.
\bibitem[{Kowalski-Glikman(2001)}]{Kowalski-Glikman:2001vvk}
\bibinfo{author}{J.~Kowalski-Glikman},
\newblock \bibinfo{title}{{Observer independent quantum of mass}},
\newblock \bibinfo{journal}{Phys. Lett. A} \bibinfo{volume}{286}
  (\bibinfo{year}{2001}) \bibinfo{pages}{391--394}.
  \DOIprefix\doi{10.1016/S0375-9601(01)00465-0}.
\bibitem[{Amelino-Camelia(2001)}]{Amelino-Camelia:2000cpa}
\bibinfo{author}{G.~Amelino-Camelia},
\newblock \bibinfo{title}{{Testable scenario for relativity with minimum
  length}},
\newblock \bibinfo{journal}{Phys. Lett. B} \bibinfo{volume}{510}
  (\bibinfo{year}{2001}) \bibinfo{pages}{255--263}.
  \DOIprefix\doi{10.1016/S0370-2693(01)00506-8}.
\bibitem[{Magueijo and Smolin(2002)}]{Magueijo:2001cr}
\bibinfo{author}{J.~Magueijo}, \bibinfo{author}{L.~Smolin},
\newblock \bibinfo{title}{{Lorentz invariance with an invariant energy scale}},
\newblock \bibinfo{journal}{Phys. Rev. Lett.} \bibinfo{volume}{88}
  (\bibinfo{year}{2002}) \bibinfo{pages}{190403}.
  \DOIprefix\doi{10.1103/PhysRevLett.88.190403}.
\bibitem[{Magueijo and Smolin(2003)}]{Magueijo:2002am}
\bibinfo{author}{J.~Magueijo}, \bibinfo{author}{L.~Smolin},
\newblock \bibinfo{title}{{Generalized Lorentz invariance with an invariant
  energy scale}},
\newblock \bibinfo{journal}{Phys. Rev. D} \bibinfo{volume}{67}
  (\bibinfo{year}{2003}) \bibinfo{pages}{044017}.
  \DOIprefix\doi{10.1103/PhysRevD.67.044017}.
\bibitem[{Mota et~al.(2019)Mota, Santos, Grams, da~Silva, and
  Menezes}]{Mota:2019zln}
\bibinfo{author}{C.~E. Mota}, \bibinfo{author}{L.~C.~N. Santos},
  \bibinfo{author}{G.~Grams}, \bibinfo{author}{F.~M. da~Silva},
  \bibinfo{author}{D.~P. Menezes},
\newblock \bibinfo{title}{{Combined Rastall and Rainbow theories of gravity
  with applications to neutron stars}},
\newblock \bibinfo{journal}{Phys. Rev. D} \bibinfo{volume}{100}
  (\bibinfo{year}{2019}) \bibinfo{pages}{024043}.
  \DOIprefix\doi{10.1103/PhysRevD.100.024043}.
\bibitem[{Debnath(2021)}]{Debnath:2019eor}
\bibinfo{author}{U.~Debnath},
\newblock \bibinfo{title}{{Charged Gravastars in Rastall-Rainbow Gravity}},
\newblock \bibinfo{journal}{Eur. Phys. J. Plus} \bibinfo{volume}{136}
  (\bibinfo{year}{2021}) \bibinfo{pages}{442}.
  \DOIprefix\doi{10.1140/epjp/s13360-021-01460-6}.
\bibitem[{Das et~al.(2022)Das, Maity, Saha, and Debnath}]{Das:2022vxq}
\bibinfo{author}{K.~P. Das}, \bibinfo{author}{S.~Maity},
  \bibinfo{author}{P.~Saha}, \bibinfo{author}{U.~Debnath},
\newblock \bibinfo{title}{{Charged anisotropic strange star in Rastall-Rainbow
  gravity}},
\newblock \bibinfo{journal}{Mod. Phys. Lett. A} \bibinfo{volume}{37}
  (\bibinfo{year}{2022}) \bibinfo{pages}{2250201}.
  \DOIprefix\doi{10.1142/S0217732322502017}.
\bibitem[{Mota et~al.(2022)Mota, Santos, da~Silva, Flores, da~Silva, and
  Menezes}]{Mota:2022zbq}
\bibinfo{author}{C.~E. Mota}, \bibinfo{author}{L.~C.~N. Santos},
  \bibinfo{author}{F.~M. da~Silva}, \bibinfo{author}{C.~V. Flores},
  \bibinfo{author}{T.~J.~N. da~Silva}, \bibinfo{author}{D.~P. Menezes},
\newblock \bibinfo{title}{{Anisotropic compact stars in
  Rastall\textendash{}Rainbow gravity}},
\newblock \bibinfo{journal}{Class. Quant. Grav.} \bibinfo{volume}{39}
  (\bibinfo{year}{2022}) \bibinfo{pages}{085008}.
  \DOIprefix\doi{10.1088/1361-6382/ac5a13}.
\bibitem[{Li et~al.(2024)Li, Yang, and Lin}]{Li:2023fux}
\bibinfo{author}{J.~Li}, \bibinfo{author}{B.~Yang}, \bibinfo{author}{W.~Lin},
\newblock \bibinfo{title}{{Massive white dwarfs in Rastall-Rainbow gravity}},
\newblock \bibinfo{journal}{JCAP} \bibinfo{volume}{04} (\bibinfo{year}{2024})
  \bibinfo{pages}{081}. \DOIprefix\doi{10.1088/1475-7516/2024/04/081}.
  \href{http://arxiv.org/abs/2305.17676}{{\tt arXiv:2305.17676}}.
\bibitem[{Tangphati et~al.(2023)Tangphati, Muniz, Pradhan, and
  Banerjee}]{Tangphati:2023nwz}
\bibinfo{author}{T.~Tangphati}, \bibinfo{author}{C.~R. Muniz},
  \bibinfo{author}{A.~Pradhan}, \bibinfo{author}{A.~Banerjee},
\newblock \bibinfo{title}{{Traversable wormholes in Rastall-Rainbow gravity}},
\newblock \bibinfo{journal}{Phys. Dark Univ.} \bibinfo{volume}{42}
  (\bibinfo{year}{2023}) \bibinfo{pages}{101364}.
  \DOIprefix\doi{10.1016/j.dark.2023.101364}.
\bibitem[{Pradhan et~al.(2024)Pradhan, Islam, Zeyauddin, and
  Banerjee}]{Pradhan:2023vhn}
\bibinfo{author}{A.~Pradhan}, \bibinfo{author}{S.~Islam},
  \bibinfo{author}{M.~Zeyauddin}, \bibinfo{author}{A.~Banerjee},
\newblock \bibinfo{title}{{Noncommutative effects on wormholes in
  Rastall\textendash{}Rainbow gravity}},
\newblock \bibinfo{journal}{Int. J. Mod. Phys. D} \bibinfo{volume}{33}
  (\bibinfo{year}{2024}) \bibinfo{pages}{2450008}.
  \DOIprefix\doi{10.1142/S0218271824500081}.
\bibitem[{Jyothilakshmi et~al.(2023)Jyothilakshmi, Naik, and
  Sreekanth}]{Jyothilakshmi:2023cao}
\bibinfo{author}{O.~P. Jyothilakshmi}, \bibinfo{author}{L.~J. Naik},
  \bibinfo{author}{V.~Sreekanth},
\newblock \bibinfo{title}{{Bose-Einstein condensate stars in combined
  Rastall-Rainbow gravity}}  (\bibinfo{year}{2023}).
  \href{http://arxiv.org/abs/2311.13813}{{\tt arXiv:2311.13813}}.
\bibitem[{Tangphati et~al.(2024)Tangphati, Gogoi, Pradhan, and
  Banerjee}]{Tangphati:2023fey}
\bibinfo{author}{T.~Tangphati}, \bibinfo{author}{D.~J. Gogoi},
  \bibinfo{author}{A.~Pradhan}, \bibinfo{author}{A.~Banerjee},
\newblock \bibinfo{title}{{Investigating stable quark stars in Rastall-Rainbow
  gravity and their compatibility with gravitational wave observations}},
\newblock \bibinfo{journal}{JHEAp} \bibinfo{volume}{42} (\bibinfo{year}{2024})
  \bibinfo{pages}{12--20}. \DOIprefix\doi{10.1016/j.jheap.2024.02.006}.
\bibitem[{Li et~al.(2024)Li, Yang, and Lin}]{Li:2024uwv}
\bibinfo{author}{J.~Li}, \bibinfo{author}{B.~Yang}, \bibinfo{author}{W.~Lin},
\newblock \bibinfo{title}{{Color-flavor locked quark stars in
  Rastall\textendash{}Rainbow gravity}},
\newblock \bibinfo{journal}{Chin. J. Phys.} \bibinfo{volume}{89}
  (\bibinfo{year}{2024}) \bibinfo{pages}{134--143}.
  \DOIprefix\doi{10.1016/j.cjph.2024.03.003}.
\bibitem[{Tolman(1930)}]{Tolman:1930zz}
\bibinfo{author}{R.~C. Tolman},
\newblock \bibinfo{title}{{On the Use of the Energy-Momentum Principle in
  General Relativity}},
\newblock \bibinfo{journal}{Phys. Rev.} \bibinfo{volume}{35}
  (\bibinfo{year}{1930}) \bibinfo{pages}{875--895}.
  \DOIprefix\doi{10.1103/PhysRev.35.875}.
\bibitem[{Szabados(2009)}]{Szabados:2009eka}
\bibinfo{author}{L.~B. Szabados},
\newblock \bibinfo{title}{{Quasi-Local Energy-Momentum and Angular Momentum in
  General Relativity}},
\newblock \bibinfo{journal}{Living Rev. Rel.} \bibinfo{volume}{12}
  (\bibinfo{year}{2009}) \bibinfo{pages}{4}.
  \DOIprefix\doi{10.12942/lrr-2009-4}.
\bibitem[{Herrera and Santos(1997)}]{Herrera:1997plx}
\bibinfo{author}{L.~Herrera}, \bibinfo{author}{N.~O. Santos},
\newblock \bibinfo{title}{{Local anisotropy in self-gravitating systems}},
\newblock \bibinfo{journal}{Phys. Rept.} \bibinfo{volume}{286}
  (\bibinfo{year}{1997}) \bibinfo{pages}{53--130}.
  \DOIprefix\doi{10.1016/S0370-1573(96)00042-7}.
\bibitem[{Bel(1961)}]{bel1961}
\bibinfo{author}{L.~Bel},
\newblock \bibinfo{title}{Inductions {\'e}lectromagn{\'e}tique et
  gravitationnelle},
\newblock \bibinfo{journal}{Ann. Inst. H Poincar{\'e}} \bibinfo{volume}{17}
  (\bibinfo{year}{1961}) \bibinfo{pages}{37--57}. \URLprefix
  \url{http://www.numdam.org/item?id=AIHP_1961__17_1_37_0}.
\bibitem[{Herrera et~al.(2009)Herrera, Ospino, Di~Prisco, Fuenmayor, and
  Troconis}]{Herrera:2009zp}
\bibinfo{author}{L.~Herrera}, \bibinfo{author}{J.~Ospino},
  \bibinfo{author}{A.~Di~Prisco}, \bibinfo{author}{E.~Fuenmayor},
  \bibinfo{author}{O.~Troconis},
\newblock \bibinfo{title}{{Structure and evolution of self-gravitating objects
  and the orthogonal splitting of the Riemann tensor}},
\newblock \bibinfo{journal}{Phys. Rev. D} \bibinfo{volume}{79}
  (\bibinfo{year}{2009}) \bibinfo{pages}{064025}.
  \DOIprefix\doi{10.1103/PhysRevD.79.064025}.
\bibitem[{Lemaitre(1933)}]{Lemaitre:1933gd}
\bibinfo{author}{G.~Lemaitre},
\newblock \bibinfo{title}{{The expanding universe}},
\newblock \bibinfo{journal}{Annales Soc. Sci. Bruxelles A} \bibinfo{volume}{53}
  (\bibinfo{year}{1933}) \bibinfo{pages}{51--85}.
  \DOIprefix\doi{10.1023/A:1018855621348}.
\bibitem[{Ruderman(1972)}]{Ruderman:1972aj}
\bibinfo{author}{M.~Ruderman},
\newblock \bibinfo{title}{{Pulsars: structure and dynamics}},
\newblock \bibinfo{journal}{Ann. Rev. Astron. Astrophys.} \bibinfo{volume}{10}
  (\bibinfo{year}{1972}) \bibinfo{pages}{427--476}.
  \DOIprefix\doi{10.1146/annurev.aa.10.090172.002235}.
\bibitem[{Hillebrandt and Steinmetz(1976)}]{Hillebrandt1976}
\bibinfo{author}{W.~Hillebrandt}, \bibinfo{author}{K.~O. Steinmetz},
\newblock \bibinfo{title}{{Anisotropic neutron star models: stability against
  radial and nonradial pulsations}},
\newblock \bibinfo{journal}{Astron. Astrophys} \bibinfo{volume}{53}
  (\bibinfo{year}{1976}) \bibinfo{pages}{283--287}.
\bibitem[{Bowers and Liang(1974)}]{Bowers:1974tgi}
\bibinfo{author}{R.~L. Bowers}, \bibinfo{author}{E.~P.~T. Liang},
\newblock \bibinfo{title}{{Anisotropic Spheres in General Relativity}},
\newblock \bibinfo{journal}{Astrophys. J.} \bibinfo{volume}{188}
  (\bibinfo{year}{1974}) \bibinfo{pages}{657--665}.
  \DOIprefix\doi{10.1086/152760}.
\bibitem[{Sokolov(1980)}]{Sokolov1980}
\bibinfo{author}{A.~I. Sokolov},
\newblock \bibinfo{title}{{Phase transitions in a superfluid neutron liquid}},
\newblock \bibinfo{journal}{Sor Phys. JETP} \bibinfo{volume}{52}
  (\bibinfo{year}{1980}).
\bibitem[{Sawyer(1972)}]{Sawyer:1972cq}
\bibinfo{author}{R.~F. Sawyer},
\newblock \bibinfo{title}{{Condensed pi- phase in neutron star matter}},
\newblock \bibinfo{journal}{Phys. Rev. Lett.} \bibinfo{volume}{29}
  (\bibinfo{year}{1972}) \bibinfo{pages}{382--385}.
  \DOIprefix\doi{10.1103/PhysRevLett.29.382}.
\bibitem[{Khan et~al.(2024)Khan, Adeel, and Yousaf}]{Khan:2024vsh}
\bibinfo{author}{S.~Khan}, \bibinfo{author}{A.~Adeel},
  \bibinfo{author}{Z.~Yousaf},
\newblock \bibinfo{title}{{Structure of anisotropic fuzzy dark matter black
  holes}},
\newblock \bibinfo{journal}{Eur. Phys. J. C} \bibinfo{volume}{84}
  (\bibinfo{year}{2024}) \bibinfo{pages}{572}.
  \DOIprefix\doi{10.1140/epjc/s10052-024-12940-1}.
\bibitem[{Mak and Harko(2003)}]{Mak:2001eb}
\bibinfo{author}{M.~K. Mak}, \bibinfo{author}{T.~Harko},
\newblock \bibinfo{title}{{Anisotropic stars in general relativity}},
\newblock \bibinfo{journal}{Proc. Roy. Soc. Lond. A} \bibinfo{volume}{459}
  (\bibinfo{year}{2003}) \bibinfo{pages}{393--408}.
  \DOIprefix\doi{10.1098/rspa.2002.1014}.
\bibitem[{Ovalle(2017)}]{Ovalle:2017fgl}
\bibinfo{author}{J.~Ovalle},
\newblock \bibinfo{title}{{Decoupling gravitational sources in general
  relativity: from perfect to anisotropic fluids}},
\newblock \bibinfo{journal}{Phys. Rev. D} \bibinfo{volume}{95}
  (\bibinfo{year}{2017}) \bibinfo{pages}{104019}.
  \DOIprefix\doi{10.1103/PhysRevD.95.104019}.
\bibitem[{Albalahi et~al.(2024{\natexlab{a}})Albalahi, Yousaf, Ali, and
  Khan}]{Albalahi:2024vpy}
\bibinfo{author}{A.~M. Albalahi}, \bibinfo{author}{Z.~Yousaf},
  \bibinfo{author}{A.~Ali}, \bibinfo{author}{S.~Khan},
\newblock \bibinfo{title}{{Isotropization and complexity shift of
  gravitationally decoupled charged anisotropic sources}},
\newblock \bibinfo{journal}{Eur. Phys. J. C} \bibinfo{volume}{84}
  (\bibinfo{year}{2024}{\natexlab{a}}) \bibinfo{pages}{9}.
  \DOIprefix\doi{10.1140/epjc/s10052-023-12358-1}.
\bibitem[{Albalahi et~al.(2024{\natexlab{b}})Albalahi, Bhatti, Ali, and
  Khan}]{Albalahi:2024ujg}
\bibinfo{author}{A.~M. Albalahi}, \bibinfo{author}{M.~Z. Bhatti},
  \bibinfo{author}{A.~Ali}, \bibinfo{author}{S.~Khan},
\newblock \bibinfo{title}{{Electromagnetic field on the complexity of minimally
  deformed compact stars}},
\newblock \bibinfo{journal}{Eur. Phys. J. C} \bibinfo{volume}{84}
  (\bibinfo{year}{2024}{\natexlab{b}}) \bibinfo{pages}{293}.
  \DOIprefix\doi{10.1140/epjc/s10052-024-12652-6}.
\bibitem[{Yousaf et~al.(2024)Yousaf, Khan, Turki, and Suzuki}]{Yousaf:2024fkr}
\bibinfo{author}{Z.~Yousaf}, \bibinfo{author}{S.~Khan}, \bibinfo{author}{N.~B.
  Turki}, \bibinfo{author}{T.~Suzuki},
\newblock \bibinfo{title}{{Modeling of self-gravitating compact configurations
  using radial metric deformation approach}},
\newblock \bibinfo{journal}{Chin. J. Phys.} \bibinfo{volume}{89}
  (\bibinfo{year}{2024}) \bibinfo{pages}{1595--1610}.
  \DOIprefix\doi{10.1016/j.cjph.2024.04.012}.
\bibitem[{Contreras and Stuchlik(2022)}]{Contreras:2022vec}
\bibinfo{author}{E.~Contreras}, \bibinfo{author}{Z.~Stuchlik},
\newblock \bibinfo{title}{{A simple protocol to construct solutions with
  vanishing complexity by Gravitational Decoupling}},
\newblock \bibinfo{journal}{Eur. Phys. J. C} \bibinfo{volume}{82}
  (\bibinfo{year}{2022}) \bibinfo{pages}{706}.
  \DOIprefix\doi{10.1140/epjc/s10052-022-10684-4}.
\bibitem[{Finch and Skea(1989)}]{Finch_1989}
\bibinfo{author}{M.~R. Finch}, \bibinfo{author}{J.~E.~F. Skea},
\newblock \bibinfo{title}{A realistic stellar model based on an ansatz of
  duorah and ray},
\newblock \bibinfo{journal}{Class. Quant. Grav.} \bibinfo{volume}{6}
  (\bibinfo{year}{1989}) \bibinfo{pages}{467}.
  \DOIprefix\doi{10.1088/0264-9381/6/4/007}.
\bibitem[{Habsi et~al.(2023)Habsi, Maurya, Badri, Al-Alawiya, Mukhaini, Malki,
  and Mustafa}]{Habsi:2023stx}
\bibinfo{author}{M.~A. Habsi}, \bibinfo{author}{S.~K. Maurya},
  \bibinfo{author}{S.~A. Badri}, \bibinfo{author}{M.~Al-Alawiya},
  \bibinfo{author}{T.~A. Mukhaini}, \bibinfo{author}{H.~A. Malki},
  \bibinfo{author}{G.~Mustafa},
\newblock \bibinfo{title}{{Self-bound embedding Class I anisotropic stars by
  gravitational decoupling within vanishing complexity factor formalism}},
\newblock \bibinfo{journal}{Eur. Phys. J. C} \bibinfo{volume}{83}
  (\bibinfo{year}{2023}) \bibinfo{pages}{286}.
  \DOIprefix\doi{10.1140/epjc/s10052-023-11420-2}.
\bibitem[{Andrade and Santana(2023)}]{Andrade:2023jvm}
\bibinfo{author}{J.~Andrade}, \bibinfo{author}{D.~Santana},
\newblock \bibinfo{title}{{An anisotropic stellar fluid configuration with
  vanishing complexity}},
\newblock \bibinfo{journal}{Eur. Phys. J. C} \bibinfo{volume}{83}
  (\bibinfo{year}{2023}) \bibinfo{pages}{523}.
  \DOIprefix\doi{10.1140/epjc/s10052-023-11701-w}.
\bibitem[{Khan and Yousaf(2024)}]{Khan:2024jnb}
\bibinfo{author}{S.~Khan}, \bibinfo{author}{Z.~Yousaf},
\newblock \bibinfo{title}{{Complexity-free charged anisotropic Finch-Skea model
  satisfying Karmarkar condition}},
\newblock \bibinfo{journal}{Phys. Scripta} \bibinfo{volume}{99}
  (\bibinfo{year}{2024}) \bibinfo{pages}{055303}.
  \DOIprefix\doi{10.1088/1402-4896/ad38e2}.
\bibitem[{Schwarzschild(1916)}]{Schwarzschild:1916ae}
\bibinfo{author}{K.~Schwarzschild},
\newblock \bibinfo{title}{{On the gravitational field of a sphere of
  incompressible fluid according to Einstein's theory}},
\newblock \bibinfo{journal}{Sitzungsber. Preuss. Akad. Wiss. Berlin (Math.
  Phys.)} \bibinfo{volume}{1916} (\bibinfo{year}{1916})
  \bibinfo{pages}{424--434}. \href{http://arxiv.org/abs/physics/9912033}{{\tt
  arXiv:physics/9912033}}.
\bibitem[{Buchdahl(1959)}]{Buchdahl:1959zz}
\bibinfo{author}{H.~A. Buchdahl},
\newblock \bibinfo{title}{{General Relativistic Fluid Spheres}},
\newblock \bibinfo{journal}{Phys. Rev.} \bibinfo{volume}{116}
  (\bibinfo{year}{1959}) \bibinfo{pages}{1027}.
  \DOIprefix\doi{10.1103/PhysRev.116.1027}.

\end{thebibliography}
\bibliographystyle{elsarticle-num-names}
\end{document}